\documentstyle[art12,bezier,emlines2]{article}
\normalbaselines
\input{tcilatex}
\begin{document}
\voffset=-2cm
\begin{center}
{\Large \bf Self-accelerated Universe}\footnote{An extended version of the
author's talk given at the {\it 9th Adriatic Meeting: Particle Physics and the 
Universe}, held in Dubrovnik, Croatia, 4--14 September 2003}\\[3mm]
{B.\ P.\ Kosyakov}\\[5mm]
{\it  Institute for Theoretical and Mathematical Physics,}\\[3mm]
{\it  Russian Federal Nuclear Center--VNIIEF, Sarov 607190, Russia}\\[3mm]
E-mail address: {\verb+kosyakov@vniief.ru+}
\end{center}
\begin{abstract}
\noindent
It is widely believed that the large redshifts for distant supernovae 
are due to the vacuum energy dominance, or, more precisely, due to a 
cosmological constant in  Einstein's equations,
which is responsible for the anti-gravitation effect.
A tacit assumption is that particles move along geodesics for the 
background metric. 
This is in the same spirit as the consensus regarding the uniform Galilean 
motion of a free electron. 
We note, however, that, apart from the Galilean solution, there is a 
self-accelerated solution 
to the Abraham--Lorentz--Dirac equation governing the behavior of a radiating 
electron. 
Likewise, a self-accelerated solution to the entire system of equations, both 
gravitation and matter equations of motion including, may exist, which 
provides an 
alternative explanation for the accelerated expansion of the Universe, 
without recourse to the hypothetic cosmological constant.
\end{abstract}
\section{Introduction}
\label
{introduction}
The recent measurements of redshifts for Type Ia supernovae \cite{Riess}
suggest that the Universe expansion is accelerating.
To interpret this discovery, one usually write the Friedmann 
equation
\begin{equation}
H^2=\left(\frac{\dot a}{a}\right)^2=\frac{2}{A_V^{2}}+\frac{2A_D}{a^{3}}+
\frac{2A_B}{a^{3}}+\frac{A_R}{a^{4}}-k
\label
{Friedmann-sol}
\end{equation}                                          
where  $H$ is the Hubble expansion parameter,  $a$ is the scale factor,
$A_V$, $A_D$, $A_B$, and $A_R$ are Friedmann integrals of the 
motion related to the energy density of vacuum, dark matter, 
nonrelativistic particles (baryons), and radiation;
$k$ is the spatial curvature, with $k=1, -1,0$ 
corresponding to the closed, open, and flat models.
Equation (\ref{Friedmann-sol})
is derived from Einstein's equations with a positive
cosmological constant
\begin{equation}
R_{\mu\nu}-\frac12\,g_{\mu\nu}\, R-\Lambda\,g_{\mu\nu}=-{8\pi G}\,T_{\mu\nu}
\label
{Einstein}
\end{equation}                                          
using the generic form of the line element for homogeneous and isotropic 
spacetimes
\begin{equation}
ds^2=dt^2- a^2 F(r)^2 d\Omega^2 -a^2 dr^2.
\label
{ds-Friedmann}
\end{equation}                                          
Here, $t$ is the proper time,  $d\Omega^2=d\theta^2+ \sin^2\theta\,d\varphi^2$,
$F=\sin r,\,\sinh r,\, r$ for $k=1,-1,0$, respectively, the 
cosmological
constant $\Lambda$ relates to the vacuum energy density $\rho_V$ 
as $\Lambda=8\pi G\rho_V$, and $A_V=(8\pi G\rho_V/3)^{-1/2}$. 
A pseudo-Riemannian space with 
arbitrary curvilinear coordinates $x^\mu$ and metric tensor $g_{\mu\nu}$ is 
understood in this cosmological model.

In the expanding Universe, the scale factor $a$ increases with time.
So,
there comes a time when the first term in (\ref{Friedmann-sol}) becomes
dominant.
The asymptotic ($t\to\infty$) solution to Eq.\ (\ref{Friedmann-sol}) is
\begin{equation}
a(t)=A_V f(t),\quad
f(t)=
\cases{\cosh(t/A_V)& $k=1$\cr
       \sinh(t/A_V)& $k=-1$\cr
       \exp(t/A_V) & $k=0$\cr},  
\label
{DE-Sitter-sol}
\end{equation}                                          
which shows that the cosmological expansion accelerates, ${\ddot a}>0$.

Thus, the presence of a positive cosmological constant $\Lambda$ in 
(\ref{Einstein}), which is responsible for the anti-gravitation effect, 
ensures the accelerated expansion of the Universe.
At present, this explanation of the large redshift data 
for distant supernovae is widely accepted (for a recent review see \cite{Padmanabhan}) 
with the tacit belief that particles (galaxies, clusters, etc.) 
move along geodesics for the background metric $g_{\mu\nu}$. 
To impeach this belief, we first remark that galaxies and clusters
every so often have intrinsic angular momentum, spin.
It is well known, however, that a spinning particle is deflected from the 
geodesic \cite{Papapetrou}--\cite{Schild}.
Note that such an anomalous motion is unrelated to the spacetime geometry:
a spinning particle in Minkowski space behaves in a non-Galilean manner.
Indeed, let us consider a classical spinning particle in the Frenkel model
\cite{frenkel}.
Its motion, in the absence of external forces, is governed 
by the equation (see, e.\ g., \cite{k02})\footnote{Throughout this paper we 
use the Minkowski metric 
with signature $-2$.
The velocity of light is taken unity.}
\begin{equation}
{\cal S}^2{\ddot v}^{\mu}+M^2v^{\mu}=m p^{\mu}
\label
{eq-motion}
\end{equation}                                          
where ${\cal S}$ is the spin magnitude, $v^{\mu}={\dot z}^{\mu}$ is the  
four-velocity, the dot denotes the derivative with respect to the proper 
time $s$,  $p^{\mu}$ is the four-momentum (which is constant for the free 
particle), $M$ and $m$ are
the mass and rest mass, defined as $M^2 =p^2$ and 
$m=p\cdot v$.
For $p^2>0$, $p^\mu=$ const, the general solution to Eq.\ (\ref{eq-motion}) is
\begin{equation}
z^{\mu}( s)=\frac{m}{M^2}\,p^{\mu} s+\frac{\alpha^{\mu}}
{\omega}\,\cos \omega s+\frac{\beta^{\mu}}{\omega}\,\sin \omega s
\label
{x-solution}
\end{equation}                                          
where $\alpha\cdot p=\beta\cdot p=\alpha\cdot\beta=0$,\, $\alpha^2=\beta^2$,
and $\omega=M/{\cal S}$.
This helical world line describes motion called the Zitterbewegung 
(this phenomenon was first discovered by Schr\"odinger \cite{Schrodinger} in 
a quantum context).
For $p^2<0$, $p^\mu=$ const, we have
\begin{equation}
z^{\mu}(s)=-\frac{m}{{\cal M}^2}\,p^{\mu} s+\frac{\alpha^{\mu}}{\Omega}\,
{\cosh}\,\Omega s+\frac{\beta^{\mu}}{\Omega}\,{\sinh}\,\Omega s
\label
{x-hyperb}
\end{equation}                                          
where ${\cal M}^2=-p^2$, $\Omega={\cal M}/{\cal S}$, and
$\alpha^2=-\beta^2$.
This solution describes motion with increasing velocity.

One may argue that spacelike four-momenta $p^\mu$ are highly unnatural for 
classical particles.
While this is a strong objection, it seems reasonable to say that both 
solutions (\ref{x-solution}) and (\ref{x-hyperb}) support the 
idea of non-Galilean regimes for free spinning particles. 

Another fact deserving of notice is that massive particles can emit 
gravitational waves.
It is conceivable that a massive particle emitting gravitational waves
moves in a runaway regime, that is, deviates sharply from a 
geodesic for the background metric.
It seems plausible that runaway solutions may offer an alternative 
explanation 
for the accelerated
expansion of the Universe, without recourse to the cosmological constant  
hypothesis.
If this explanation is true, one would gain the most benefit from it looking at
a theoretical framework where $\Lambda=0$, e.\ g., unbroken supersymmetry, 
for clues of the mystery of the cosmological constant \cite{Padmanabhan}.

It is interesting to compare a massive particle emitting  gravitational waves 
and a charged particle emitting electromagnetic waves.
The nonrelativistic equation of motion for a classical electron, 
called the Abraham--Lorentz equation 
(see, e.\ g., \cite{Jackson}--\cite{L-L}), 
\begin{equation}
m {\bf a}-\frac23\,e^2\,\frac{d{\bf a}}{dt}={\bf f},
\label
{Abraham-Lorentz}
\end{equation}                        
in the absence of external forces ${\bf f}=0$, becomes
\begin{equation}
{\bf a}-\tau_0\,\frac{d{\bf a}}{dt}=0
\label
{Abraham-Lorentz-free}
\end{equation}       
where
\begin{equation} 
\tau _{0}={2e^2\over 3m}\approx 6\cdot 10^{-24}\,{\rm s}.
\label
{class-radius}
\end{equation}                        %
The general solution to  Eq.\ (\ref{Abraham-Lorentz-free}),
\begin{equation}
{\bf a}(t)={\bf A}\,\exp(t/\tau_0),
\label
{runaway-3D}
\end{equation}       
where ${\bf A}$ is the initial acceleration at $t=0$, describes
runaway (or self-accelerated) motion.
For ${\bf A}=0$, we have ${\bf a}=0$, and ${\bf v}=$ const.
Thus a free electron can behave as both Galilean 
(${\bf A}=0$), and non-Galilean (${\bf A}\ne 0$) objects.

Where does the Abraham--Lorentz equation come from?
The scheme of its derivation is as follows.
We first solve Maxwell's equations 
\begin{equation}
\Box A^{\mu}(x)=4\pi  e\int_{-\infty}^\infty\! ds\,v^\mu(s)\,
\delta^4\bigl(x-z(s)\bigr) 
\label
{Maxw-tens-2nd}
\end{equation}           
where the world line of a single charge $z^\mu(s)$ is taken to be 
arbitrarily prescribed timelike smooth curve.
The retarded Lien\'ard--Wiechert solution
\begin{equation}
A^\mu(x) =\frac{e\,v^\mu}{(x-z)\cdot v}\Bigl|_{s=s_{\rm ret}}
\label
{A-ret}
\end{equation}                                             
is regularized 
and
substituted into the equation of motion for a bare charged particle
\begin{equation}
m_0 {\bf a}=e\left({\bf E}+{{\bf v}}\times{\bf B}\right),
\label
{Lorentz}
\end{equation}                        
where $m_0$ is the bare mass.
We then require that the renormalized mass 
\begin{equation}
m=\lim_{\epsilon\to 0}\left(m_0(\epsilon)+  
{e^2\over 2\epsilon}\right)
\label
{m-ren}
\end{equation}                        
be a finite positive quantity.
Finally, we  arrive
at the Abraham--Lorentz equation (\ref{Abraham-Lorentz}) in the limit of the 
regularization removal $\epsilon\to 0$.

In order to derive the equation of motion for a massive particle capable of 
emitting gravitational waves, one should repeat the essentials of this 
procedure: find the retarded solution to Einstein's equations (\ref{Einstein}) 
with $\Lambda=0$ assuming that a given point
particle which generates the retarded gravitational field moves along an 
arbitrarily prescribed 
timelike smooth world line, regularize this solution, substitute it
into the equation of motion for the bare particle, perform the mass 
renormalization, and 
remove the regularization.
This will yield the desired equation of motion for a dressed massive particle,
which is apparently different from the equation of a geodesic 
\begin{equation}
\frac{dv^\lambda}{ds}+\Gamma^\lambda_{\mu\nu}v^\mu v^\nu=0
\label
{geodesics}
\end{equation}
where $\Gamma_{\lambda\mu\nu}$ is the Christoffel symbol for
the 
background metric $g_{\mu\nu}$ 
\begin{equation}
\Gamma_{\lambda\mu\nu}=\frac12\left(
\partial_\lambda g_{\mu\nu}-\partial_\mu  g_{\nu\lambda}-
\partial_\nu  g_{\lambda\mu}\right).
\label
{christoffel-lower}
\end{equation}

This project is highly nontrivial.
Even the first stage of it has defied implementation: by now, 
no retarded solution to the gravitation equations (\ref{Einstein}) similar to 
the Lien\'ard--Wiechert solution (\ref{A-ret}) in electrodynamics is found.
There is in addition a conceptual problem.
While the kinematical criterion for availability of electromagnetic 
radiation is that the world line of the source is curved (which makes absolute
geometric 
sense in the Maxwell--Lorentz theory formulated in Minkowski space),
such a criterion is unsound for the gravitational field source, say, a 
field singularity in a smooth manifold, since all timelike world lines of 
this point source are regarded as diffeomorphic-equivalent in the framework of general 
relativity.  
A plausible resolution of this problem is to discriminate between 
geodesics for a given background metric and world lines deviating from 
these geodesics.
It may be disappointing that we can not offer a complete explanation 
for the accelerated expansion of the Universe based on self-accelerated 
solutions analogous to solution (\ref{runaway-3D}) of the Abraham--Lorentz 
equation.
Nevertheless, relying on results drawn from solvable 
theories in which particles interact with scalar, Yang--Mills, and 
linearized gravitational fields, together with dimensional considerations, we 
cope with a modest task: guess some essential features of 
the behavior of a particle emitting gravitation waves.

The paper is organized as follows.
Section 2 demonstrates that, regardless of warnings, the self-accelerated 
solution 
(\ref{runaway-3D}) is consistent with every fundamental physical principle
and experimental fact. 
In the next section, we take a brief look at 
the self-deceleration and the equations of motion for dressed particles 
interacting 
with the  Yang--Mills, scalar, and tensor fields.
In the final section we sketch the broad outline of the equation of 
motion for a dressed massive particle interacting with gravitation field and
its relevance to the accelerated regime of the Universe development. 

\section{Dressed charged particles}
Many people, being prejudiced against solution (\ref{runaway-3D}),
accuse it of two 
sins: violation of energy conservation and lack of 
experimental evidence for self-accelerated motions. 
Our immediate task is to show that both accusations are unjust. 

The generalization of the Abraham--Lorentz equation 
(\ref{Abraham-Lorentz}) to the special relativistic context is 
the Abraham--Lorentz--Dirac equation 
(see, e.\ g., \cite{Jackson}--\cite{L-L})
\begin{equation}
ma^\lambda-{2\over 3}\,e^2\bigl({\dot a}^\lambda +v^\lambda a^2 \bigr)
=f^{\lambda}
\label
{LD}
\end{equation}                        
where $a^\mu={\dot v}^\mu$ is the four-acceleration.
This equation accounts for the dynamics of a synthesized object
whose inertia is characterized by the
quantity $m$ defined in Eq.\ (\ref{m-ren}) containing both mechanical and 
electromagnetic contributions.
We will call this object the dressed charged particle.
The state of the dressed particle is specified by the four-coordinate
of the singular field point $z^\mu$ and the four-momentum
\begin{equation}
p^\mu= m\,v^\mu-{{2\over 3}}\,e^2 a^\mu
\label
{p-mu-ren-em}
\end{equation}                        
assigned to this point.
(Teitelboim \cite{Teitelboim} was the first to derive this expression 
with the aid of some invariant regularization of the retarded
Lien\'ard--Wiechert field.)  
The singular field point is governed by the equation
\begin{equation}
\stackrel{\scriptstyle v}{\bot}({\dot p}-f)=0
\label
{Eq-Newton}
\end{equation}                        
where $\stackrel{\scriptstyle v}{\bot}$ is the operator
\begin{equation}
\stackrel{\scriptstyle v}{\bot}\,=\, 
{\bf 1}-\frac{v\otimes v}{v^2}
\label
{projector-def}
\end{equation}                        
projecting vectors on the hyperplane perpendicular to the world line,
$p^\mu$ is the four-momentum defined in (\ref{p-mu-ren-em}),
and $f^\mu$ is an external four-force applied to the point $z^\mu$.
Indeed, substitution of (\ref{p-mu-ren-em}) in (\ref{Eq-Newton}) gives the
Abraham--Lorentz--Dirac equation (\ref{LD}).
On the other hand, Eq.\ (\ref{Eq-Newton}) is nothing but Newton's second law 
in a coordinate-free form \cite{k02}.

The Abraham--Lorentz equation 
(\ref{Abraham-Lorentz}) may also be cast as Newton's second law in its
original form
\begin{equation}
\frac{d{\bf p}}{dt}={\bf f}
\label
{newton's-second-law}
\end{equation}                        
where ${\bf p}$ is the three-momentum of a nonrelativistic dressed particle,
\begin{equation}
{\bf p}= m\,{\bf v}-{{2\over 3}}\,e^2 {\bf a}\,,
\label
{p-ren-em}
\end{equation}                        %
which is obtained from the space component of Eq.\ (\ref{p-mu-ren-em}) in the  
limit 
${\bf v}\to 0$.

Teitelboim \cite{Teitelboim} was able to show that the Abraham--Lorentz--Dirac 
equation (\ref{LD}) is equivalent to the local energy-momentum balance,
\begin{equation}
{\dot p}^\mu+{\dot{\cal P}}^\mu+{\dot{\cal{\wp}}}^\mu_{}
=0,
\label
{loc-mom-balance}
\end{equation}                        
where the four-momentum of the dressed particle $p^\mu$ is defined in
(\ref{p-mu-ren-em}),  the four-momentum rate of radiation emitted by the charge
${\dot{\cal P}}^\mu$ is represented by the Larmor formula,
\begin{equation}
{\dot{\cal P}}^\mu
=-{{2\over 3}}\,e^2\,v^\mu a^2,
\label
{Larmor}
\end{equation}                        
and the external four-momentum rate ${\dot{\wp}}^\mu$ relates to the external 
Lorentz four-force,
\begin{equation}
{\dot{\wp}}^\mu=-f^\mu.
\label
{Lorentz-ex}
\end{equation}                        
The balance equation (\ref{loc-mom-balance}) reads: the four-momentum 
extracted from the external field $d{\wp}^{\mu}=-f^\mu ds$  is spent on 
the variation of the four-momentum
of the dressed particle $dp^{\mu}$ and the four-momentum  ${d{{\cal P}}}^\mu$
carried away by the radiation.

With $f^\mu=0$,
Eq.\ (\ref{LD}) is satisfied by
\begin{equation}
v^\mu( s)=\alpha^\mu{\cosh}(w_0\tau_0 e^{s/\tau_0})+ \beta^\mu{\sinh}%
(w_0\tau_0 e^{s/\tau_0})  
\label
{run-away}
\end{equation}
where $\alpha^\mu$ and $\beta^\mu$ are constant four-vectors that meet the
conditions
$\alpha\cdot\beta=0,\, \alpha^2=-\beta^2=1$,
$w_0$ is an initial acceleration magnitude, and $\tau_0$ is given by
(\ref{class-radius}).
The solution (\ref{run-away}) describes a runaway
motion, which degenerates to the Galilean regime when $w_0=0$.

We see from (\ref{runaway-3D}) and (\ref{run-away}) that the class of Galilean
world lines is distinct from the class of runaway world lines. 
A dressed particle may either show itself constantly as a Galilean object or 
execute perpetually a self-accelerated motion.
It is impossible to render a Galilean dressed particles self-accelerated
and vice versa.
The non-Galilean behavior is an innate feature of some species of dressed 
particles.
(It is remarkable that the Maxwell--Lorentz theory alone gives no way of 
deducing the distribution of dressed charged particles by these two classes.
In fact, the class of self-accelerated dressed particles may well be empty 
for some unknown reason.)

This observation makes it clear that the mere existence of runaways does not 
contradict to 
Newton's first law (`every particle continues in its state 
of rest or uniform motion in a straight line unless it is acted upon by some 
exterior force') since the essence of this law is to forbid 
the Galilean regime from changing into another regime at a finite instant.  

It is often claimed that the solution (\ref{run-away}) is nonphysical since 
the picture where a free particle continually accelerates and  continually
radiates seems contrary to energy conservation.
(Notice that a `particle' possessing the four-momentum 
$p^{\mu}=mv^{\mu }$, with its time component $p^0=m\gamma$ being positive 
definite, is usually meant in this claim.)
The energy of both this `particle' and the electromagnetic field increases for no 
apparent reason.

It is the idea of a `particle' with  such a four-momentum which 
is deceitful and opens up numerous puzzles and paradoxes \cite{k92}.
Indeed, while on the subject of what is governed by the
Abraham--Lorentz--Dirac equation (\ref{LD}), one usually imagines an aggregate composed 
of a point charge and a 
Coulomb-like field train dragged behind it.
The ascription of the four-momentum $p^{\mu}=mv^{\mu}$ to this aggregate
lacks support from a careful
analysis based on the appropriate definition of the radiation and invariant 
regularization procedures \cite{Teitelboim}. 

By contrast, proceeding from the concept of a dressed particle with the 
four-momentum defined in (\ref{p-mu-ren-em}), we have the balance equation 
(\ref{loc-mom-balance}). 
Hence, there
is no contradiction with energy conservation: 
in the absence of
external fields, the energy 
variation of the dressed particle $dp^{0}$ is equal to the energy carried away
by the radiation $-{\dot{{\cal P}}}_{{}}^{0}ds$. 
A subtlety is that the energy of the dressed particle
\begin{equation}
p^{0}=m\gamma \,\bigl(1-\tau _{0}\,\gamma ^{3}\,{\bf a}\cdot {\bf v}\bigr)
\label{p-0-em}
\end{equation}                                           
is not positive definite. 
The indefiniteness of expression (\ref{p-0-em}) means that the increase of 
velocity  need not be accompanied by the increase of energy.
It would, therefore, make no sense to inquire from where the particle extracts 
energy to accelerate itself.
The energy of the self-accelerated dressed particle is actually diminished.

The fact that the energy of a dressed particle is indefinite might appear at
first sight strange and counter-intuitive.
But it is scarcely surprising if we recall the 
synthetic origin of the dressed particle, and note that the two 
contributions to $m$, defined in (\ref{m-ren}), are  opposite 
in sign.

If $m=0$, which is another reasonable option of the mass renormalization 
(\ref{m-ren}), then the first term of Eq.\ (\ref{LD}) disappears, and,
with $f^\mu=0$, it reduces to   
\[(\stackrel{\scriptstyle v}{\bot}{\dot a})^{\hskip0.3mm\mu}=0.\]
This is the equation of relativistic uniformly accelerated motion
\cite{Rohrlich}.
Therefore, the world line of a dressed particle with $m=0$ 
in the absence of external 
forces is  a hyperbola
\begin{equation}
v^\mu(s)={\alpha^\mu}{}\,{\cosh}\,w_0 s
+{\beta^\mu}{}\,{\sinh}\,w_0 s,
\quad\alpha\cdot\beta=0,\quad \alpha^2=-\beta^2=1.
\label
{hyperb-motion}
\end{equation}                          
The constant curvature $w_0$ of such a world line may be arbitrary, in
particular $w_0=0$. 

It follows from (\ref{p-mu-ren-em}) that 
\begin{equation}
M^2={p^2}=m^2\,\bigl({1+\tau_0^2\,a^2}\bigr).
\label
{M-m}
\end{equation}                          
If $\tau_0^2\,a^2<-1$, then the dressed particle
turns to the tachyonic state, 
by which is meant a  state with $p^2<0$.
(We point out that our consideration is restricted to timelike smooth 
world lines, and the tachyonic state under  discussion is unrelated to superluminal
motions; more precisely, the advent of spacelike four-momenta is peculiar to 
sufficiently large curvatures
of the world line, rather than spacelike run of it.)

Let a dressed particle be moving in the runaway regime   
(\ref{runaway-3D}).
Then, after a lapse of time
\begin{equation}
\Delta t=-{\tau_0}\,\log\,\tau_0\bigl|{\bf A}\bigr|,
\label
{Delta-s}
\end{equation}                          
the critical acceleration $\vert{\bf a}\vert=\tau_0^{-1}$ will be exceeded, 
and the four-momentum of this dressed particle will be spacelike.

This result provides an explanation for the fact that self-accelerated dressed 
charged particles, if any, were never observed.
The period of time over which a self-accelerated electron possesses 
timelike four-momenta is quite tiny.
From (\ref{Delta-s})  and (\ref{class-radius}), the period $\Delta t$
is estimated at $\tau_0\sim 10^{-23}$ s 
for electrons, and still shorter for more massive charged elementary particles.
All primordial self-accelerated particles with such $\tau_0$'s have long 
been in the tachyonic state.
However, we have not slightest notion of how tachyons can be experimentally 
recorded.
(It seems plausible that self-accelerated particles, transmuted into tachyons, 
represent part of dark matter.) 

If a cosmological object is considered as a dressed particle emitting 
gravitational waves,
the characteristic period  $\tau_0$ may be found to be comparable with the 
inverse current Hubble scale $H^{-1}$.
The self-acceleration
of such an object can indeed be observed at the present time.
For clarity, the experimental value of this scale is 
$H^{-1}=(46\pm 4)\cdot 10^{16}$ s.

\section{Self-deceleration}
Consider a dressed colored particle (`quark' for short) in the cold QCD phase \cite{k8}.
The emission of Yang-Mills waves by an accelerated quark in this phase is 
attended with energy gains, rather than energy losses.
Accordingly, the equation of motion for a dressed quark with the color charge
$Q$ in an external Yang--Mills field $F^{\mu\nu}$
is
\begin{equation}
m\,\bigl[a^\mu+\ell\,\bigl({\dot a}^\mu +v^\mu a^2 \bigr)\bigr]=
{\rm tr}(Q F^{\mu\nu})v_\nu
\label
{eq-motion-YM-part}
\end{equation}                        
where $m$ is the renormalized mass, $\ell$ is the characteristic period,
\begin{equation}
\ell=\frac{8}{3mg^2}\left(1-\frac{1}{{\cal N}}\right),
\label{tau-0-YM}
\end{equation}
$g$ is the Yang--Mills coupling constant, and ${\cal N}\ge 2$ is the number of 
colors.

Although both equations (\ref{eq-motion-YM-part})
and (\ref{LD}) contain 
the so-called Abraham term
\begin{equation} 
\Gamma^\mu={\dot a}^\mu+v^\mu a^2,
\label
{Abraham}
\end{equation}                        %
they show marked distinction in that
$\Gamma$ is multiplied by coefficients of opposite sign.

For $F^{\mu\nu}=0$, the general solution to
Eq.\ (\ref{eq-motion-YM-part}) 
\begin{equation}
v^\mu( s)=\alpha^\mu{\cosh}(w_0\ell\,e^{- s/\ell})+
\beta^\mu{\sinh}(w_0\ell\,e^{- s/\ell}),
\label
{self-disseleration}
\end{equation}                        
with $\alpha\cdot\beta=0,\, \alpha^2=-\beta^2=1$,
describes self-decelerated motion.

By contrast, in the hot QCD phase (where the Yang--Mills--Wong system is 
Abelian, and hence linearized \cite{k8}), 
the equation of motion for a dressed quark 
is similar to (\ref{LD}), and, therefore, dressed quarks execute self-accelerated
motions like those given by (\ref{run-away}).

At first sight, the self-deceleration is quite innocuous
phenomenon, because the motion becomes almost indistinguishable from
Galilean in the short run.
However, the presence of self-decelerations does jeopardize
the consistency of the theory.
Indeed, as we go to the past, the acceleration increases exponentially, and
the rate of the energy gain grows along with it.
Thus, the energy of the Yang--Mills field at any finite instant is divergent.
On the other hand, the self-acceleration does not play such a fatal role in
electrodynamics with the retarded boundary condition (and also the hot QCD phase);
this non-Galilean regime of motion does not entail `infrared' divergences. 
Indeed, substitution of (\ref{run-away}) in (\ref{Larmor}) shows that the 
electromagnetic field energy radiated by a self-accelerated particle during
a half-infinite period of time is finite.

A plausible resolution of the trouble with `infrared' divergences due to
a non-Galilean evolution of dressed quarks in the Yang--Mills 
theory\footnote{The above conclusion regarding the 
existence of two impermeable classes of dressed particles, Galilean and non-Galilean, holds for
the present discussion.
These classes are populated according to requirements which are 
beyond the control of the Yang--Mills--Wong theory.
It is possible that some yet-to-be-known requirement for the class of
self-decelerated quarks is that it be empty.} is to 
impose the supplementary condition that the acceleration of dressed quarks be 
always less than the critical one, $\vert a \vert\le \ell^{-1}$, in other 
words, the curvature of the allowable world lines must not exceed $\ell^{-1}$.
If, for some reason, the critical
acceleration is yet attained in the cold phase, then the dressed quark 
must plunge into the hot phase, rather than turn to the tachyonic state. 
Thus, the interconversion from one phase to another is a means to 
circumvent the difficulty with infinite energy storage in some nonlinear 
field theories.

Let us briefly run through the non-Galilean behavior of dressed particles 
interacting with massless scalar and tensor  fields (for a review of the
Lagrangian description of bare particles coupled with these fields see, e.\ g., 
\cite{Barut}).
The equation of motion for a dressed particle 
in an external scalar field 
$\phi$ is \cite{Barut75}
\begin{equation}
\frac{d}{ds}\,(m+g\phi)\,v^\mu-\frac{1}{3}\,g^2({\dot a}^\mu+a^2 v^\mu)=
g\,\partial^\mu\phi
\label
{eq-motion-scalar}
\end{equation}                        
where $m$ is the renormalized mass, and $g$ is the coupling constant.
The coefficient of the Abraham term in this equation is of the same sign as 
that in (\ref{LD}), and so the non-Galilean behavior for the 
object governed by 
Eq.\ (\ref{eq-motion-scalar}) can be classified as self-accelerated  motion.

There are several versions of the interaction of a dressed particle with a 
massless symmetric tensor field $\phi_{\alpha\beta}$ \cite{Barut,Barut75}, 
among which one, being a combination of interactions with scalar and tensor 
fields known as the `linearized gravity', holds the greatest interest for
our discussion. 
The equation of motion for a dressed particle  in 
an external tensor field $\phi_{\alpha\beta}$ of the linearized gravity with the
retarded boundary condition is \cite{Havas,Barut75}
\[
\frac{d}{ds}\left[v^\mu\left(1-\frac12\,v^\alpha v^\beta\phi_{\alpha\beta} -
\frac12\,\phi^\alpha_{\hskip0.5mm\alpha}\right)+
v_\alpha \phi^{\alpha\mu}\right]+\frac{11}{3}\,G m({\dot a}^\mu+a^2 v^\mu)
\]
\begin{equation}
= \frac12\, v^\alpha v^\beta \partial^\mu\phi_{\alpha\beta}-\frac14\,
\partial^\mu\phi^\alpha_{\hskip0.5mm\alpha}
\label
{eq-motion-lin-grav}
\end{equation}                        
where $G$ is the gravitational constant, and $m$ is the renormalized mass.
The coefficient of the Abraham term in this equation is of the same sign as 
that in (\ref{eq-motion-YM-part}), which implies that the non-Galilean regime for this dressed
particle refers to self-decelerated  motion.
While on the subject of the linearized gravity in Minkowski
space as a field theory on its own right, not an approximation to general 
relativity, it should presumably be thought of as inconsistent, because the 
system is linear (unlike the Yang--Mills system), and, therefore,  
reveals a single phase.

The above results concerning the equations 
of motion for different dressed particles
are summarized in the following table
\begin{center}
\begin{tabular}{|c||c||c||c|}  
\multicolumn{4}{c}{Table 1.\quad{Abraham term in different theories}}\\[2mm]\hline 
{Scalar field} &{Abelian}     &{Yang--Mills field}  & {Linearized}  \\ 
{}             &{vector field}&{(cold phase)} & {gravity}        \\ \hline\hline
{$-\frac13\,g^2\Gamma^\mu$}&{$-\frac23\,e^2\Gamma^\mu$}&{$\frac23\,|Q^2|\Gamma^\mu$}
&{$\frac{11}{3}\,Gm^2\Gamma^\mu$}\\
&              &              & {}  \\\hline
\end{tabular}
\end{center}

\section{Discussion}
Everybody, who is confident of the physical reality of gravitational 
radiation\footnote{Infeld
\cite{Infeld} argued that gravitational radiation does not occur at all.
However, his consideration is based on the approximation method
of Einstein, Infeld, and Hoffmann, invoking peculiar assumptions, and his 
findings, being sensitive to a particular choice of coordinate and
boundary conditions, are open to argument.}, should recognize 
that the equation of motion for a massive particle emitting 
gravitational waves is different from 
the equation of a geodesic (\ref{geodesics}).
For dimensional reasons, the desired equation must include, apart from terms 
of equation (\ref{geodesics}), some higher-derivative terms, a 
diffeomorphic-covariant 
generalization of the 
Abraham term\footnote{Conceivably, this generalization is given by a 
complicated integro-differential
expression similar to that for a dressed charged particle in a curved space 
\cite{BrehmedeWitt}.}, which would result in an additional
solution distinct from that describing the geodesic for the background 
metric\footnote{Again, one may expect the existence of two disjoint classes of
dressed particles, geodesic and non-geodesic.
It may well be that some unknown fundamental principle, unrelated to the 
gravitation theory, 
requires that the class of non-geodesic particles be empty.
We assume that such is not the case.}.

As discussed in Section 3, the coefficient of the Abraham term in 
the linearized gravity is positive, hence a massive dressed particle
seems to move in a self-decelerated fashion.
This result is adverse to the suggested explanation for
the accelerated expansion of the Universe.

Remember, however, that the linarized gravity is not a 
well-defined approximation to general relativity. 
In fact, there is no unambiguous approximation scheme which is, in its 
physical outcome, independent of coordinate and boundary conditions.
For example, the famous Einstein--Infeld--Hoffmann approximation operates on
the premise that the time derivative of any field quantity is much 
smaller than the spatial derivatives, which has proved itself in nonrelativistic 
problems.
With this approximation, the gravitational dipole moment of 
a dressed particle about 
the center-of-mass vanishes  (together with its gravitational radiation), and 
the 
derivatives of acceleration  in the resulting equation of motion 
are cancelled, leaving room for higher order derivatives.

An alternative approximation procedure \cite{Havas}, which treats time and 
space coordinates on the same footing, and utilize the retarded boundary 
condition, results in the Lorentz-invariant 
equation of motion for a dressed particle (\ref{eq-motion-lin-grav}). 
However,
terms of the same order as the Abraham term may appear implicitly 
in (\ref{eq-motion-lin-grav}) through the retardation effect in the metric;
they are of opposite sign and cancel the main contribution to the
Abraham term. 

It is well to bear in mind that the runaway solution 
(\ref{runaway-3D}) contains $\exp(3mt/2e^2)$, that is, reveals an essential 
singularity at $e=0$.
This implies that perturbative treatments of Einstein's equations, using 
`smallness' of the gravitation coupling constant $G$, may be found to be
unsuitable for the 
analysis of nonperturbative effects, non-geodesic motions in particular.

On the other hand, there is good indirect evidence
that the pulsar PSR $1913+16$, which is a 
constituent of a binary neutron star system, loses (rather than gains) energy 
for its gravitational emission, rendering the observed rotation period of this 
system gradually decreasing \cite{Taylor}.  
This result lends support to our assumption that the actual deviation from a
geodesic is represented by a self-acceleration.

Let us estimate the characteristic time 
$\tau_0=\frac{11}{3}Gm$ in (\ref{eq-motion-lin-grav}).
Taking $m$ to be a typical 
cluster mass, we have $\tau_0\sim 10^8\, {\rm s}\sim 3$ years, which is 
very far from the desired value $\tau_0\sim H^{-1}\sim 10^{10}$ years.

One may hope that some way out would be found if, e.\ g., the interactions with 
other fields (electromagnetic, dilaton, gluon, etc.) would be 
appropriately combined with gravitational interaction.
Contributions of different signs from these interactions
may have a dramatic effect on the
coefficient of the aggregate Abraham term.
In addition, one should keep in mind that the representations of galaxies and 
clusters as simple poles of the gravitational field may be not adequate, 
since these objects are endowed 
with intrinsic angular momentum. 
However, the description of a radiating charged particle with spin is a challenging 
problem (for a review see \cite{Rowe}), discarded here, let alone the description of a 
spinning massive particle emitting gravitational waves.

However, the most striking possibility resides in the fact that  the characteristic time 
$\tau_0={Gm}$ with $m$ being the total visible mass of the Universe
is of order of the inverse current Hubble scale $H^{-1}$. 
Does this mean that the Universe as a whole executes self-accelerated motion?
Should this be the case, then a pronounced anisotropy in the velocity 
distribution of distant supernovae would be an observable consequence. 

\vskip3mm
\noindent
{\bf Acknowledgment}

I would like to thank I.\ D.\ Novikov for helpful comments.
This work was supported in part by the International Science and Technology 
Center under the Project {\#} 840.

\end{document}